\documentclass[a4paper,fleqn,usenatbib]{mnras}

\usepackage{newtxtext,newtxmath}

\usepackage[T1]{fontenc}
\usepackage{ae,aecompl}

\usepackage{graphicx}	
\usepackage{amsmath}	
\usepackage{amssymb}	

\title[Can we constrain the aftermath of BNS mergers?]{Can we constrain the aftermath of binary neutron star mergers with short gamma-ray bursts?}

\author[B. Patricelli et al.]{
B. Patricelli$^{1,2}$\thanks{E-mail: barbara.patricelli@pi.infn.it}
and M.G. Bernardini$^{3}$
\\
$^{1}$ Physics Department, University of Pisa, Largo B. Pontecorvo 3, I-56127 Pisa, Italy\\
$^{2}$ INFN - Sezione di Pisa, Largo B. Pontecorvo 3, I-56127 Pisa, Italy\\
$^{3}$ INAF - Osservatorio Astronomico di Brera, via Bianchi 46, I-23807 Merate (LC), Italy\\
}

\date{Accepted XXX. Received YYY; in original form ZZZ}

\pubyear{2020}

\begin{document}
\label{firstpage}
\pagerange{\pageref{firstpage}--\pageref{lastpage}}
\maketitle

\begin{abstract}
The joint observation of GW170817 and GRB170817A proved that binary neutron star (BNS) mergers are progenitors of short Gamma-ray Bursts (SGRB): this established a direct link between the still unsettled  SGRB central engine and the outcome of BNS mergers, whose nature depends on the equation of state (EOS) and on the masses of the NSs. We propose a novel method to probe the central engine of SGRBs based on this link. We produce an extended catalog of BNS mergers by combining recent theoretically predicted BNS merger rate as a function of redshift and the NS mass distribution inferred from measurements of Galactic BNSs. We use this catalog to predict the number of BNS systems ending as magnetars (stable or Supramassive NS) or BHs (formed promptly or after the collapse of a hypermassive NS) for different EOSs, and we compare these outcomes with the observed rate of SGRBs. Despite the uncertainties mainly related to the poor knowledge of the SGRB jet structure, we find that for most EOSs the rate of magnetars produced after BNS mergers is sufficient to power all the SGRBs, while scenarios with only BHs as possible central engine seems to be disfavoured.

\end{abstract}




\section{Introduction}

The discovery of GW 170817, the first Binary Neutron Star (BNS) merger event in the domain of Gravitational Waves (GWs; \citealp{2017PhRvL.119p1101A}), and its association with the Short Gamma-Ray Burst (SGRB) 170817A (\citealp{2017ApJ...848L..13A,2017ApJ...848L..14G,2017ApJ...848L..15S}) provided the first direct proof that BNS mergers are indeed progenitors of SGRBs. However, the nature of the central engine powering SGRBs is still unsettled.

The outcome of a BNS merger mainly depends on the masses of the two compact objects and on the Equation of State (EOS) of nuclear matter (see, e.g., \citealp{2006PhRvD..73f4027S,2008PhRvD..78h4033B,2011PhRvD..83l4008H}). Specifically, the BNS merger can form: i) a Black Hole (BH) from prompt collapse; ii) a hypermassive Neutron Star (HMNS), that collapse into a BH in a very short time scale, of the order of ms to $\sim$ 100 ms; iii) a supramassive NS (SMNS), that collapse into a BH on a time scale of the order of seconds, minutes or longer and iv) a stable NS. 

SMNSs and stable NSs endowed with large magnetic field and short spin period at birth (magnetars) are thought to be the central object powering GRBs \citep{1998A&A...333L..87D,2001ApJ...552L..35Z,2009ApJ...702.1171C,2011MNRAS.413.2031M}. This proposal has been revealed itself very successful in reproducing the observed properties of the subclass of GRBs with X-ray plateau \citep{2013MNRAS.430.1061R} and/or SGRBs with extended emission (EE, \citealp{2008MNRAS.385.1455M,2012MNRAS.419.1537B}), at least from a phenomenological point of view (see e.g. \citealp{2015JHEAp...7...64B}, and \citealp{2020arXiv200701124L} for some recent results). Nevertheless, we are still lacking of a direct proof for the magnetar central engine. Alternative explanations for the observed X-ray plateau have been recently proposed in the framework of the structured-jet model, without any assumption on the nature of the central engine \citep{2020MNRAS.492.2847B,2020ApJ...893...88O}. In addition, there are several theoretical limitations for the magnetar central engine, especially related to the difficulty for the magnetar to launch an ultra-relativistic jet \citep{2020MNRAS.495L..66C}. 

For GRB 170817A, it was not possible to clearly establish the nature of the aftermath of the merger from the GW signal (see \citealp{2017ApJ...851L..16A,2019PhRvX...9a1001A,2019ApJ...875..160A}; see also the other confirmed BNS merger detected by LIGO and Virgo, GW 190425, described in  \citealp{2020ApJ...892L...3A}). The analysis of the electromagnetic (EM) emission from GRB 170817A seems to support the formation of a BH after a stage of HMNS lasting $\sim 100$ ms  (\citealp{2017ApJ...850L..19M,2017PhRvD..96l3012S,2018ApJ...852L..29R};  see however \citealp{2019MNRAS.483.1912P}). Anyway, this single case cannot exclude that magnetars may be the central engine of at least a fraction of SGRBs.

In this Letter we provide a new piece of evidence to approach this issue by investigating under which conditions the rate of magnetars (and BHs) produced in BNS mergers is sufficient to power SGRBs. We produce an extended catalog of BNS mergers covering a volume comparable to the one of detection of SGRBs by combining the theoretically predicted BNS merger rate as a function of redshift and the NS mass distribution  (Sec. \ref{sec:cosmicrate}). We use this catalog to predict the rates of the different outcomes of the BNS merger for different EOSs and, comparing them to the rate of SGRBs in the same volume (Sec. \ref{sec:ratesGRBs}), we discuss what is the most plausible scenario for the SGRB central engine (Sec. \ref{sec:results}).

\section{The BNS merging system catalog}\label{sec:cosmicrate}
We generate samples of synthetic BNS merging systems populating the Universe up to a redshift z=1, accordingly to a theoretically predicted cosmic BNS merger rate (see Sec. \ref{sec:cosmicrate}). We assign to each NS in the simulated BNS systems a mass, randomly extracted from a given mass distribution  (see Sec. \ref{sec:mass}). The outcome of the simulated BNS mergers (that, as already said, can be a NS or a BH) is then estimated according to the masses of the two NSs and for different EOSs (see Sec. \ref{sec:EOS}).
In order to have enough statistics, we perform simulations corresponding to a total observing time of 20000 years. 

\subsection{The cosmic BNS merger rate}

In this work we use the cosmic merger rate density of BNSs ($R_{\rm BNS}$) estimated in \cite{2018MNRAS.479.4391M} (hereafter M18); specifically, among the various models proposed we use the one predicting a local merger rate of 591 Gpc$^{-3}$ yr$^{-1}$, that is consistent with current estimates obtained after the LIGO-Virgo detection of GW170817 and GW190425 \citep{2020ApJ...892L...3A}.

To estimate the rate of events within $z\leq 1$, we proceed as follows. We assume standard flat $\Lambda$CDM cosmology with H$_0$= 70.4 km s$^{-1}$ Mpc$^{-1}$, $\rm{\Omega_M}$=0.2726,  $\Omega_\Lambda$=0.7274, $\rm{\Omega_K}$=0, as in \cite{2018MNRAS.479.4391M}. The comoving volume element is 
\begin{equation}
dV(z)=\frac{c}{H_0} \frac{D_c^2}{E(z)} d\Omega dz,  
\end{equation}
where c is the speed of light, $d\Omega$ is the solid angle,  \mbox{$E(z)=\sqrt{\Omega_{\rm{M}} (1+z)^3+\Omega_{\rm{K}}  (1+z)^2+\Omega_\Lambda}$}  and $D_c$ is the comoving distance, given by
\begin{equation}
D_c(z)=\frac{c}{H_0} \int_0^z{\frac{dz'}{E(z')}}    
\end{equation}
The total BNS merger rate within a redshift z in the observer frame is then given by:
\begin{equation}
n_{\rm obs}(< z)=4 \pi \int_0^z{\frac{R_{\rm BNS}(z')}{(1+z')}\frac{dV(z')}{dz'}dz'} \quad [yr^{-1}].
\end{equation}

\subsection{The NS mass distribution}\label{sec:mass}
The masses of many NSs have been estimated in the past through both EM and GW observations. 
The smallest, precisely measured NS mass through EM observations is \mbox{M=1.174 $\pm$ 0.004 M$_\odot$}, as estimated for the companion of the pulsar J0453+1559 \citep{2015ApJ...812..143M}. The most massive NS yet observed through EM  observations is MSP J0740+6620, whose mass has been estimated to be M=$2.14^{+0.10}_{-0.09}$ (68 \% credible interval,  \citealp{2020NatAs...4...72C}). 
The masses of the NSs estimated from the GW detections of BNS mergers GW170817 \citep{2019PhRvX...9a1001A} and GW190425 \citep{2020ApJ...892L...3A} are within this range when the spins are restricted to be within the range observed in Galactic BNSs. However, for GW190425 a higher mass for the one of the two NS is allowed when higher spins are considered (the mass of the primary component is in the range 1.61-2.52 M$_\odot$, see \citealp{2020ApJ...892L...3A}). 

In this work we assume that both components of the binary systems  have a mass distribution equal to the distribution inferred from measurements related to binary systems in the Galaxy: a Gaussian function with central mass 1.33 M$_\odot$ and dispersion of 0.09 M$_\odot$ \citep{2016ARA&A..54..401O}; the two gaussians are assumed to be uncorrelated. We also assume that the mass distribution does not change with redshift (see, e.g., \citealp{2019MNRAS.487....2M,2019MNRAS.482..870E}).

\subsection{The EOS and the BNS merger outcome}\label{sec:EOS}

As already discussed, besides the NS initial masses (see Sec. \ref{sec:mass}), the other ingredient needed to determine which is the BNS merger remnant is the EOS. The EOS of cold, ultra-dense matter is still poorly constrained  at  high  densities, despite  decades  of  study (see, e.g., \citealp{2001ApJ...550..426L}). A powerful instrument to better understand the properties of matter in the most extreme conditions, and therefore the EOS, is represented by GWs. Specifically, the most prominent effect of matter during the observed binary inspiral comes from the tidal deformation that each star's gravitational field induces on its companion. For instance, the  observation of GW170817 showed that ``soft'' EOSs such as APR4 \citep{1998PhRvC..58.1804A}, which predict smaller values of the tidal deformability parameter, are favored over ``stiff'' EOSs such as H4 \citep{1991PhRvL..67.2414G} and MS1 \citep{1996NuPhA.606..508M}, which predict larger values of the tidal deformability parameter and lie outside the 90\% credible region  \citep{2018PhRvL.121p1101A}. Other investigations have been done with the joint analysis of GW and EM data associated with GW170817 (see, e.g.,  \citealp{2017ApJ...850L..19M,2018ApJ...852L..29R}). Finally, further constraints on the EOS have been placed with a recent joint analysis of the NICER measurements of the mass and the radius of PSR J0030+0451 and of the GW data from GW170817 (see \citealp{2020ApJ...893L..21R}).

To evaluate how the assumed EOS affects our estimates, in this work we use three different EOSs: APR4, MS1 and H4. These three EOSs cover a relatively wide range of maximum NS masses, but all with a maximum gravitational mass $\gtrsim$ 2 M$_\odot$, consistent with the current observational EM limits. A more comprehensive set of EOSs will be considered in future studies.

To estimate the merger outcome for the various combinations of NS masses and EOS, we follow the approach  proposed by \cite{2017ApJ...844L..19P}. Specifically, we assume a reference mass lost from the system during the merger of M$_{\rm{lost}}$=0.01  M$_\odot$, we then convert the gravitational masses  of the simulated NS to baryonic masses (m$_{\rm{b,1}}$ and m$_{\rm{b,2}}$) using the model developed by \cite{2017ApJ...844L..19P} for non-rotating NS\footnote{As pointed out in \cite{2017ApJ...844L..19P}, during the inspiral phase the two NSs are not strongly affected by tidal coupling, so they are not spun appreciably and their structure is well approximated by non-spinning NS models.} and for the three different EOSs; finally, we estimate the total barionic mass of the remnant as:
\begin{equation}
\rm{M}_{b,tot}=\rm{m}_{b,1}+\rm{m}_{b,2}-\rm{M_{lost}}.
\end{equation}
The value of M$_{\rm{b,tot}}$ is then compared with the maximum NS barionic masses to evaluate the nature of the remnant (see Table 1 of \citealp{2017ApJ...844L..19P}). Remnants that are stable NSs or a SMNSs are considered as ``magnetar-like'' GRB central engine.

\section{The rate of Short Gamma-Ray Bursts}\label{sec:ratesGRBs}

The rate of SGRBs can be computed from their luminosity function and the redshift distribution. The luminosity function is usually modelled as a single or broken power law, derived by fitting the peak flux distribution of SGRBs detected by BATSE or Fermi/GBM \citep{2005A&A...435..421G,2006A&A...453..823G,2006ApJ...650..281N,2006ApJ...643L..91H,2008MNRAS.388L...6S,2011ApJ...727..109V,2014MNRAS.442.2342D,2015MNRAS.448.3026W,2016A&A...594A..84G}. The redshift distribution is assumed to follow the cosmic star formation rate with a delay due to the time necessary for the progenitor binary system to merge, and compared with that of the few SGRBs with measured $z$.

\begin{figure*}
    \includegraphics[width=0.5\textwidth]{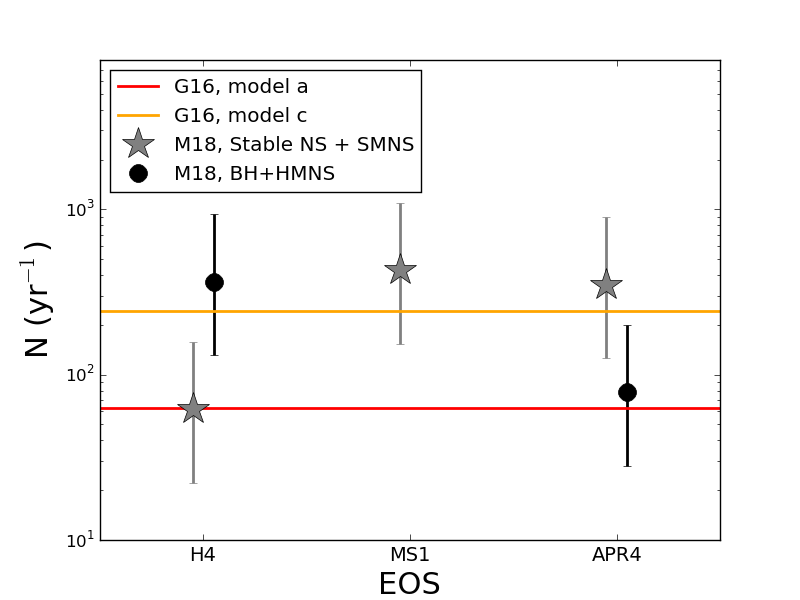}\includegraphics[width=0.5\textwidth]{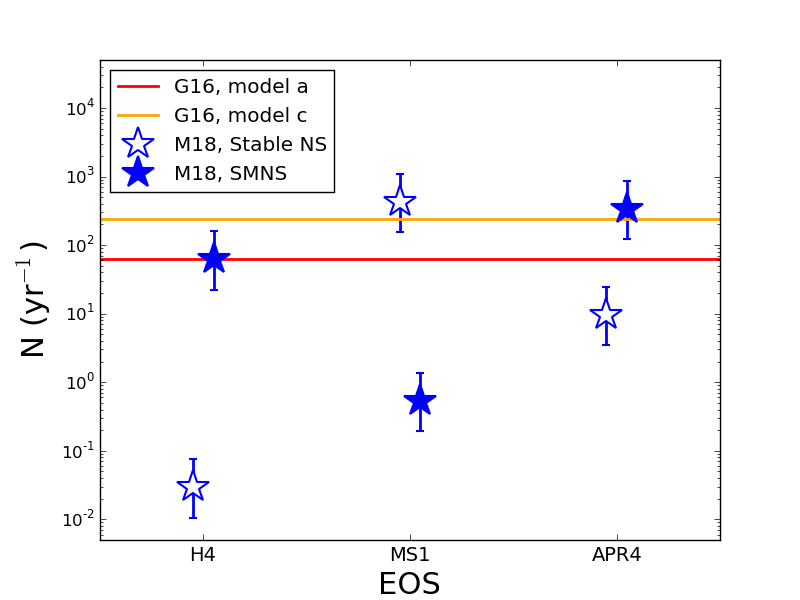}
    \caption{Event rates within redshift z=1. The solid red and orange lines represent the rate of SGRB, as estimated from the models ``a'' and ``c'' of G16  respectively. Left panel: the star (circle) represents the merger rate of BNS systems  with $\theta_i \leq \theta_j=5^\circ$ (see Sec. \ref{sec:ratesGRBs}) ending as a stable NS or a SMNS (HMNS or BH) for different EOSs, estimated using the M18 BNS cosmic merger rate; the error bars represent the merger rate interval corresponding to the assumed range of values of $\theta_j$: 3$^\circ$ - 8$^\circ$ (see Sec. \ref{sec:ratesGRBs}). Right panel: the empty (filled) blue star represents the merger rate of BNS systems ending as a stable NS (SMNS) for different EOSs, estimated using the M18 BNS cosmic merger rate.
    }
    \label{fig:VolRates}
\end{figure*}

In this work we adopt the SGRB rate derived by \cite{2016A&A...594A..84G} (hereafter G16), that is the most robust up to date because it is derived using all the available observer-frame constraints (i.e. peak flux, fluence, peak energy and duration distributions) of the large population of Fermi/GBM SGRBs and the rest-frame properties of a complete sample of SGRBs detected by Swift \citep{2014MNRAS.442.2342D}. In particular we use the rates derived under two possible assumptions: that the intrinsic ${\rm E_p - L_{iso}}$ and ${\rm E_p - E_{iso}}$ correlations hold (case ``a'' in G16, red solid line in fig.~\ref{fig:VolRates}), or that the distributions of intrinsic peak energy, luminosity, and duration are independent (case ``c'' in G16, orange solid line in fig.~\ref{fig:VolRates}). 

In order to directly compare these rates to the ones for BNS mergers derived in the previous Section, one should account for the fact that SGRB emission is collimated. Thus, we assign  to each BNS system a random inclination of the orbital plane with respect to the line of sight, and we assume that the GRB jet axis is perpendicular to the plane of the binary’s orbit (i.e., that the angle of the observer with respect to the jet is equal to the inclination angle of the BNS system). We draw from our sample the BNS mergers whose inclination angle is $\theta_i \leq \theta_j$, with $\theta_j$ the GRB jet opening angle. The jet opening angle is poorly constrained for SGRBs since the weakness of the afterglow makes a clear detection of a jet break challenging. The few estimates available range from $3^\circ$ to $8^\circ$ \citep[see e.g.][and references therein]{2014ApJ...780..118F}. Recently, \citet{2019Sci...363..968G} performed high spatial resolution measurements of the source size and displacement of the SGRB 170817A with Very Long Baseline Interferometry observations, favoring a structured jet model: a successful jet with a structured angular velocity and energy profile, featuring a narrow core with opening angle ${\rm \theta_c =3.4\pm1^\circ}$. This jet structure might be a common feature of SGRBs (quasi-universal, \citealp{2015MNRAS.450.3549S}). In this scenario the luminosity function of SGRBs would be dominated at low luminosity by events seen with large viewing angles that are detectable in the local Universe \citep[see e.g.][]{2020A&A...636A.105S}. However the rates derived by G16 correspond to SGRBs that are seen within (or close to) the inner core \citep{2020A&A...636A.105S}. We thus assume $\theta_j$ to be in the range 3$^\circ$ - 8$^\circ$, with a reference value of 5$^\circ$. 

Within the total number of SGRBs, it is possible that only a fraction of them are powered by magnetars. The magnetar central engine has been advocated to interpret phenomenologically observational features related only to specific sub-classes of SGRBs: those with an X-ray plateau, and those with an Extended Emission (EE). The first sub-class comprises those SGRBs that exhibit a flattening in the X-ray afterglow between $100 - 10^4$ s after the main event, possibly caused by the spin-down radiation of the magnetar that is expected to emit a relativistic wind at timescales comparable to the observed ones \citep{1998A&A...333L..87D,2001ApJ...552L..35Z,2009ApJ...702.1171C,2011MNRAS.413.2031M}. The presence of a plateau phase is a feature common to $\sim 50\%$ of SGRBs \citep{2013MNRAS.430.1061R,2014MNRAS.442.2342D}. The other sub-class is characterised by having a re-brightening in the prompt emission after the initial spike lasting up to $\sim 100$ s \citep{2006ApJ...643..266N}, observed in $\sim 15\%$ of cases \citep{2014ARA&A..52...43B}. This long-lasting emission can be explained in the context of the magnetar central engine as produced by either a relativistic wind powered by the magnetar rotational energy  \citep{2008MNRAS.385.1455M,2012MNRAS.419.1537B,2013ApJ...776L..40Y,2014MNRAS.439.3916M,2016ApJ...819...14S,2016ApJ...819...15S}, or by a magnetic ``propeller'' that ejects the material from the accretion disc surrounding the newly-formed magnetar \citep{2014MNRAS.438..240G}.


\section{Results and Discussion}\label{sec:results}

In this work we aim at assessing if the magnetars (either stable NSs or SMNSs) produced by BNS mergers are sufficient to power at least a fraction of SGRBs. We thus compare the observer-frame BNS merger rate giving rise to a magnetar, for the different EOSs and within redshift z=1, with the total rate of SGRBs in the same volume, as derived in G16 under two different assumptions (model ``a'' and ``c''). The results are shown in fig. \ref{fig:VolRates}, left panel. 

The percentage of BNS mergers ending as a magnetar changes according to the assumed EOS: it is $\sim 14 \%$, $\sim 82 \%$ and $\sim 100 \%$ for the H4, APR4 and MS1, respectively\footnote{These figures are slightly different with respect to the ones presented in \cite{2017ApJ...844L..19P} because of the different mass distribution used in this work.}. 
The total rates obtained with APR4 and MS1 are consistent with the total rate of SGRBs for the range of beaming angles considered (see Sec. \ref{sec:ratesGRBs}), implying that in principle all the SGRBs can be powered by magnetars. The H4 EOS predicts a rate of magnetars that is consistent with model ``a'' of G16, while is only marginally consistent with model ``c'' if we assume larger values of $\theta_j \sim 10^\circ$. 
The total number of newly born magnetars produced in BNS mergers is dominated by SMNSs for the APR4 and H4 EOSs (see fig. \ref{fig:VolRates}, right panel). Thus if we want to explain the X-ray plateau or the EE with the magnetar model, we need to require that in at least a fraction of SGRBs ($\sim 50\%$) the SMNS survives for long enough ($\sim$ minutes to hours) after the merger. This is not the case for MS1, where the number of stable NSs is sufficient to power all SGRBs (see fig. \ref{fig:VolRates}, right panel).
 
As an opposite case, we also compare the rate of BHs and HMNSs (BH+HMNS) produced after the merger of on-axis BNS systems for the different EOSs to the total rate of SGRBs (see fig. \ref{fig:VolRates}, left panel).  We group these two cases because in both scenarios the SGRB is thought to be powered by the BH formed either promptly or after the rapid collapse of the HMNS and they are distinguishable only with complementary observations (GWs and kilonova emission), as it happened for GRB 170817A \citep{2017ApJ...850L..19M,2017PhRvD..96l3012S,2018ApJ...852L..29R}. 
Only for the EOS H4 the rate of BH+HMNSs is high enough to account for all the SGRBs. For the EOS APR4, the rate of BH+HMNSs is consistent with the rate of SGRBs predicted by model ``a'' of G16, but is consistent with the rate predicted by model ``c'' of G16 only assuming larger $\theta_j \sim 9^\circ$. For the EOS MS1 the rate of merging BNS systems ending as a BH or an HMNS is negligible. This indicates that, at least for the APR4 and MS1 EOSs, BH as the only possible SGRB central engine is even disfavoured.

These results only consider the remnants of merging BNS systems formed through isolated binary evolution, that are thought to give the dominant contribution to the total BNS merger rates (see, e.g., \citealp{2020ApJ...888L..10Y,2020arXiv200409533S}).  The contribution from systems formed through dynamical assembly is not expected to change the conclusions of this paper; a detailed investigation on this will be performed in future works.  
We don't include in our study BNS merging systems as massive as the progenitor of GW190425. The formation of such heavy BNS systems, not yet observed through EM waves, is challenging for current population synthesis models (see, e.g., \citealp{2020ApJ...900...13S} and references therein); future GW observations will be crucial to confirm their existence\footnote{The possibility that one or both components of GW190425 are BHs cannot be ruled out \citep{2020ApJ...892L...3A}.}, to understand which is their formation mechanism, to better constrain the associated merger rate and also to complement the mass distribution of Galactic BNSs: this could potentially modify the results here presented, possibly leading to an increase in the rate of BH remnants.

Another possible progenitor system for SGRBs that leads to the formation of a BH is the coalescence of a NS-BH system. Since the NS-BH mergers have a very different range of mass ratios with respect to the NS-NS case, SGRBs produced via this channel should have distinct properties compared to the sub-class produced via NS-NS merger. However no conclusive dichotomy is found so far to support the production of an important fraction of SGRBs via this progenitor channel \citep{2020ApJ...895...58G}, though SGRBs with EE have been tentatively associated to NS-BH progenitors \citep{2008MNRAS.385L..10T,2020ApJ...895...58G}. In addition, the rate of NS-BH systems could also be much smaller than the one of BNSs (e.g. \citealp{2020A&A...636A.104B}), especially those with mass ratios such that the NS is disrupted and not swallowed whole (e.g. \citealp{2014PhRvD..90b4026F,2017CQGra..34d4002F}). The LIGO and Virgo collaborations did not report any confirmed detection of NS-BH mergers so far\footnote{We are not considering GW190814, whose nature is uncertain \citep{2020ApJ...896L..44A}. }, and they put an upper limit on the NS-BH merger rate of 610 Gpc$^{-3}$ yr$^{-1}$  (see \citealp{2019PhRvX...9c1040A}). For these reasons, neglecting the contribution to the rate of SGRBs from NS-BH progenitors does not affect our conclusions about the magnetar central engine, while this progenitor type provides a complementary way to produce BH central engine.

While with this work we have clearly shown that the rate of magnetars produced in BNS mergers is high enough to power SGRBs, from fig.  \ref{fig:VolRates} it is evident that the our current knowledge of the SGRB rate does not allow us to distinguish among the different EOSs. The major source of uncertainty is the poor constraints that we have on the jet opening angle due to the faintness of their late-time afterglows that makes the detection of jet-breaks challenging.  Dedicated campaigns to monitor late-time evolution of the afterglows are needed to put constraints on the jet structure: this will result also, as a by-product, in a more robust estimate of the true event rate that might be used to constrain the EOSs.
Future missions such as Athena will be key instruments to this scope  \citep{2013arXiv1306.2336J}. In addition to this in the next years, when current generation GW detectors will operate with increased sensitivity, we expect that more BNS mergers will be detected, possibly with EM counterparts: this will allow us to put tighter constraints on the merger rates of these systems and, jointly to a deeper understanding of the EOS, to better probe the magnetar-SGRB connection. The possibility to detect post-merger signals with the third-generation GW detectors as the Einstein Telescope \citep{2017JApA...38...58A} will be the ultimate proof to unambiguously identify the central engine of SGRBs.

\section*{Acknowledgements}
The authors thank Michela Mapelli for providing the cosmic BNS merger rate density used in this work, and Riccardo Ciolfi, Paolo D'Avanzo, Giancarlo Ghirlanda and  Anthony Piro for useful discussions. The authors also thank the anonymous referee for his/her valuable comments and suggestions, which helped us to improve this work. 

\section*{Data Availability}
The data underlying this article will be shared on reasonable request to the corresponding author.


\bibliographystyle{mnras}
\bibliography{Bibliography} 

\newcommand{\noop}[1]{}
\begin{thebibliography}{}
\makeatletter
\relax
\def\mn@urlcharsother{\let\do\@makeother \do\$\do\&\do\#\do\^\do\_\do\%\do\~}
\def\mn@doi{\begingroup\mn@urlcharsother \@ifnextchar [ {\mn@doi@}
  {\mn@doi@[]}}
\def\mn@doi@[#1]#2{\def\@tempa{#1}\ifx\@tempa\@empty \href
  {http://dx.doi.org/#2} {doi:#2}\else \href {http://dx.doi.org/#2} {#1}\fi
  \endgroup}
\def\mn@eprint#1#2{\mn@eprint@#1:#2::\@nil}
\def\mn@eprint@arXiv#1{\href {http://arxiv.org/abs/#1} {{\tt arXiv:#1}}}
\def\mn@eprint@dblp#1{\href {http://dblp.uni-trier.de/rec/bibtex/#1.xml}
  {dblp:#1}}
\def\mn@eprint@#1:#2:#3:#4\@nil{\def\@tempa {#1}\def\@tempb {#2}\def\@tempc
  {#3}\ifx \@tempc \@empty \let \@tempc \@tempb \let \@tempb \@tempa \fi \ifx
  \@tempb \@empty \def\@tempb {arXiv}\fi \@ifundefined
  {mn@eprint@\@tempb}{\@tempb:\@tempc}{\expandafter \expandafter \csname
  mn@eprint@\@tempb\endcsname \expandafter{\@tempc}}}

\bibitem[\protect\citeauthoryear{{Abbott} et~al.}{{Abbott}
  et~al.}{2017a}]{2017PhRvL.119p1101A}
{Abbott} B.~P.,  et~al., 2017a, \mn@doi [\prl]
  {10.1103/PhysRevLett.119.161101}, \href
  {https://ui.adsabs.harvard.edu/abs/2017PhRvL.119p1101A} {119, 161101}

\bibitem[\protect\citeauthoryear{{Abbott} et~al.}{{Abbott}
  et~al.}{2017b}]{2017ApJ...848L..13A}
{Abbott} B.~P.,  et~al., 2017b, \mn@doi [\apjl] {10.3847/2041-8213/aa920c},
  \href {https://ui.adsabs.harvard.edu/abs/2017ApJ...848L..13A} {848, L13}

\bibitem[\protect\citeauthoryear{{Abbott} et~al.}{{Abbott}
  et~al.}{2017c}]{2017ApJ...851L..16A}
{Abbott} B.~P.,  et~al., 2017c, \mn@doi [\apjl] {10.3847/2041-8213/aa9a35},
  \href {https://ui.adsabs.harvard.edu/abs/2017ApJ...851L..16A} {851, L16}

\bibitem[\protect\citeauthoryear{{Abbott} et~al.}{{Abbott}
  et~al.}{2018}]{2018PhRvL.121p1101A}
{Abbott} B.~P.,  et~al., 2018, \mn@doi [\prl] {10.1103/PhysRevLett.121.161101},
  \href {https://ui.adsabs.harvard.edu/abs/2018PhRvL.121p1101A} {121, 161101}

\bibitem[\protect\citeauthoryear{{Abbott} et~al.}{{Abbott}
  et~al.}{2019a}]{2019PhRvX...9a1001A}
{Abbott} B.~P.,  et~al., 2019a, \mn@doi [Physical Review X]
  {10.1103/PhysRevX.9.011001}, \href
  {https://ui.adsabs.harvard.edu/abs/2019PhRvX...9a1001A} {9, 011001}

\bibitem[\protect\citeauthoryear{{Abbott} et~al.}{{Abbott}
  et~al.}{2019b}]{2019PhRvX...9c1040A}
{Abbott} B.~P.,  et~al., 2019b, \mn@doi [Physical Review X]
  {10.1103/PhysRevX.9.031040}, \href
  {https://ui.adsabs.harvard.edu/abs/2019PhRvX...9c1040A} {9, 031040}

\bibitem[\protect\citeauthoryear{{Abbott} et~al.}{{Abbott}
  et~al.}{2019c}]{2019ApJ...875..160A}
{Abbott} B.~P.,  et~al., 2019c, \mn@doi [\apj] {10.3847/1538-4357/ab0f3d},
  \href {https://ui.adsabs.harvard.edu/abs/2019ApJ...875..160A} {875, 160}

\bibitem[\protect\citeauthoryear{{Abbott} et~al.}{{Abbott}
  et~al.}{2020a}]{2020ApJ...892L...3A}
{Abbott} B.~P.,  et~al., 2020a, \mn@doi [\apjl] {10.3847/2041-8213/ab75f5},
  \href {https://ui.adsabs.harvard.edu/abs/2020ApJ...892L...3A} {892, L3}

\bibitem[\protect\citeauthoryear{{Abbott} et~al.}{{Abbott}
  et~al.}{2020b}]{2020ApJ...896L..44A}
{Abbott} B.~P.,  et~al., 2020b, \mn@doi [\apjl] {10.3847/2041-8213/ab960f},
  \href {https://ui.adsabs.harvard.edu/abs/2020ApJ...896L..44A} {896, L44}

\bibitem[\protect\citeauthoryear{{Akmal}, {Pandharipande}  \&
  {Ravenhall}}{{Akmal} et~al.}{1998}]{1998PhRvC..58.1804A}
{Akmal} A.,  {Pandharipande} V.~R.,   {Ravenhall} D.~G.,  1998, \mn@doi [\prc]
  {10.1103/PhysRevC.58.1804}, \href
  {https://ui.adsabs.harvard.edu/abs/1998PhRvC..58.1804A} {58, 1804}

\bibitem[\protect\citeauthoryear{{Andersson}}{{Andersson}}{2017}]{2017JApA...38...58A}
{Andersson} N.,  2017, \mn@doi [Journal of Astrophysics and Astronomy]
  {10.1007/s12036-017-9463-8}, \href
  {https://ui.adsabs.harvard.edu/abs/2017JApA...38...58A} {38, 58}

\bibitem[\protect\citeauthoryear{{Baiotti}, {Giacomazzo}  \&
  {Rezzolla}}{{Baiotti} et~al.}{2008}]{2008PhRvD..78h4033B}
{Baiotti} L.,  {Giacomazzo} B.,   {Rezzolla} L.,  2008, \mn@doi [\prd]
  {10.1103/PhysRevD.78.084033}, \href
  {https://ui.adsabs.harvard.edu/abs/2008PhRvD..78h4033B} {78, 084033}

\bibitem[\protect\citeauthoryear{{Belczynski} et~al.,}{{Belczynski}
  et~al.}{2020}]{2020A&A...636A.104B}
{Belczynski} K.,  et~al., 2020, \mn@doi [\aap] {10.1051/0004-6361/201936528},
  \href {https://ui.adsabs.harvard.edu/abs/2020A&A...636A.104B} {636, A104}

\bibitem[\protect\citeauthoryear{{Beniamini}, {Duque}, {Daigne}  \&
  {Mochkovitch}}{{Beniamini} et~al.}{2020}]{2020MNRAS.492.2847B}
{Beniamini} P.,  {Duque} R.,  {Daigne} F.,   {Mochkovitch} R.,  2020, \mn@doi
  [\mnras] {10.1093/mnras/staa070}, \href
  {https://ui.adsabs.harvard.edu/abs/2020MNRAS.492.2847B} {492, 2847}

\bibitem[\protect\citeauthoryear{{Berger}}{{Berger}}{2014}]{2014ARA&A..52...43B}
{Berger} E.,  2014, \mn@doi [ARA\&A] {10.1146/annurev-astro-081913-035926},
  \href {http://adsabs.harvard.edu/abs/2014ARA%26A..52...43B} {52, 43}

\bibitem[\protect\citeauthoryear{{Bernardini}}{{Bernardini}}{2015}]{2015JHEAp...7...64B}
{Bernardini} M.~G.,  2015, \mn@doi [Journal of High Energy Astrophysics]
  {10.1016/j.jheap.2015.05.003}, \href
  {https://ui.adsabs.harvard.edu/abs/2015JHEAp...7...64B} {7, 64}

\bibitem[\protect\citeauthoryear{{Bucciantini}, {Metzger}, {Thompson}  \&
  {Quataert}}{{Bucciantini} et~al.}{2012}]{2012MNRAS.419.1537B}
{Bucciantini} N.,  {Metzger} B.~D.,  {Thompson} T.~A.,   {Quataert} E.,  2012,
  \mn@doi [MNRAS] {10.1111/j.1365-2966.2011.19810.x}, \href
  {http://adsabs.harvard.edu/abs/2012MNRAS.419.1537B} {419, 1537}

\bibitem[\protect\citeauthoryear{{Ciolfi}}{{Ciolfi}}{2020}]{2020MNRAS.495L..66C}
{Ciolfi} R.,  2020, \mn@doi [\mnras] {10.1093/mnrasl/slaa062}, \href
  {https://ui.adsabs.harvard.edu/abs/2020MNRAS.495L..66C} {495, L66}

\bibitem[\protect\citeauthoryear{{Corsi} \& {M{\'e}sz{\'a}ros}}{{Corsi} \&
  {M{\'e}sz{\'a}ros}}{2009}]{2009ApJ...702.1171C}
{Corsi} A.,  {M{\'e}sz{\'a}ros} P.,  2009, \mn@doi [ApJ]
  {10.1088/0004-637X/702/2/1171}, \href
  {http://adsabs.harvard.edu/abs/2009ApJ...702.1171C} {702, 1171}

\bibitem[\protect\citeauthoryear{{Cromartie} et~al.,}{{Cromartie}
  et~al.}{2020}]{2020NatAs...4...72C}
{Cromartie} H.~T.,  et~al., 2020, \mn@doi [Nature Astronomy]
  {10.1038/s41550-019-0880-2}, \href
  {https://ui.adsabs.harvard.edu/abs/2020NatAs...4...72C} {4, 72}

\bibitem[\protect\citeauthoryear{{D'Avanzo} et~al.,}{{D'Avanzo}
  et~al.}{2014}]{2014MNRAS.442.2342D}
{D'Avanzo} P.,  et~al., 2014, \mn@doi [\mnras] {10.1093/mnras/stu994}, \href
  {https://ui.adsabs.harvard.edu/abs/2014MNRAS.442.2342D} {442, 2342}

\bibitem[\protect\citeauthoryear{{Dai} \& {Lu}}{{Dai} \&
  {Lu}}{1998}]{1998A&A...333L..87D}
{Dai} Z.~G.,  {Lu} T.,  1998, A\&A, \href
  {http://adsabs.harvard.edu/abs/1998A%26A...333L..87D} {333, L87}

\bibitem[\protect\citeauthoryear{{Eldridge}, {Stanway}  \& {Tang}}{{Eldridge}
  et~al.}{2019}]{2019MNRAS.482..870E}
{Eldridge} J.~J.,  {Stanway} E.~R.,   {Tang} P.~N.,  2019, \mn@doi [\mnras]
  {10.1093/mnras/sty2714}, \href
  {https://ui.adsabs.harvard.edu/abs/2019MNRAS.482..870E} {482, 870}

\bibitem[\protect\citeauthoryear{{Fong} et~al.,}{{Fong}
  et~al.}{2014}]{2014ApJ...780..118F}
{Fong} W.,  et~al., 2014, \mn@doi [\apj] {10.1088/0004-637X/780/2/118}, \href
  {https://ui.adsabs.harvard.edu/abs/2014ApJ...780..118F} {780, 118}

\bibitem[\protect\citeauthoryear{{Foucart} et~al.,}{{Foucart}
  et~al.}{2014}]{2014PhRvD..90b4026F}
{Foucart} F.,  et~al., 2014, \mn@doi [\prd] {10.1103/PhysRevD.90.024026}, \href
  {https://ui.adsabs.harvard.edu/abs/2014PhRvD..90b4026F} {90, 024026}

\bibitem[\protect\citeauthoryear{{Foucart} et~al.,}{{Foucart}
  et~al.}{2017}]{2017CQGra..34d4002F}
{Foucart} F.,  et~al., 2017, \mn@doi [Classical and Quantum Gravity]
  {10.1088/1361-6382/aa573b}, \href
  {https://ui.adsabs.harvard.edu/abs/2017CQGra..34d4002F} {34, 044002}

\bibitem[\protect\citeauthoryear{{Ghirlanda} et~al.,}{{Ghirlanda}
  et~al.}{2016}]{2016A&A...594A..84G}
{Ghirlanda} G.,  et~al., 2016, \mn@doi [\aap] {10.1051/0004-6361/201628993},
  \href {https://ui.adsabs.harvard.edu/abs/2016A&A...594A..84G} {594, A84}

\bibitem[\protect\citeauthoryear{{Ghirlanda} et~al.,}{{Ghirlanda}
  et~al.}{2019}]{2019Sci...363..968G}
{Ghirlanda} G.,  et~al., 2019, \mn@doi [Science] {10.1126/science.aau8815},
  \href {https://ui.adsabs.harvard.edu/abs/2019Sci...363..968G} {363, 968}

\bibitem[\protect\citeauthoryear{{Glendenning} \& {Moszkowski}}{{Glendenning}
  \& {Moszkowski}}{1991}]{1991PhRvL..67.2414G}
{Glendenning} N.~K.,  {Moszkowski} S.~A.,  1991, \mn@doi [\prl]
  {10.1103/PhysRevLett.67.2414}, \href
  {https://ui.adsabs.harvard.edu/abs/1991PhRvL..67.2414G} {67, 2414}

\bibitem[\protect\citeauthoryear{{Goldstein} et~al.,}{{Goldstein}
  et~al.}{2017}]{2017ApJ...848L..14G}
{Goldstein} A.,  et~al., 2017, \mn@doi [\apjl] {10.3847/2041-8213/aa8f41},
  \href {https://ui.adsabs.harvard.edu/abs/2017ApJ...848L..14G} {848, L14}

\bibitem[\protect\citeauthoryear{{Gompertz}, {O'Brien}  \& {Wynn}}{{Gompertz}
  et~al.}{2014}]{2014MNRAS.438..240G}
{Gompertz} B.~P.,  {O'Brien} P.~T.,   {Wynn} G.~A.,  2014, \mn@doi [MNRAS]
  {10.1093/mnras/stt2165}, \href
  {http://adsabs.harvard.edu/abs/2014MNRAS.438..240G} {438, 240}

\bibitem[\protect\citeauthoryear{{Gompertz}, {Levan}  \& {Tanvir}}{{Gompertz}
  et~al.}{2020}]{2020ApJ...895...58G}
{Gompertz} B.~P.,  {Levan} A.~J.,   {Tanvir} N.~R.,  2020, \mn@doi [\apj]
  {10.3847/1538-4357/ab8d24}, \href
  {https://ui.adsabs.harvard.edu/abs/2020ApJ...895...58G} {895, 58}

\bibitem[\protect\citeauthoryear{{Guetta} \& {Piran}}{{Guetta} \&
  {Piran}}{2005}]{2005A&A...435..421G}
{Guetta} D.,  {Piran} T.,  2005, \mn@doi [\aap] {10.1051/0004-6361:20041702},
  \href {https://ui.adsabs.harvard.edu/abs/2005A&A...435..421G} {435, 421}

\bibitem[\protect\citeauthoryear{{Guetta} \& {Piran}}{{Guetta} \&
  {Piran}}{2006}]{2006A&A...453..823G}
{Guetta} D.,  {Piran} T.,  2006, \mn@doi [\aap] {10.1051/0004-6361:20054498},
  \href {https://ui.adsabs.harvard.edu/abs/2006A&A...453..823G} {453, 823}

\bibitem[\protect\citeauthoryear{{Hopman}, {Guetta}, {Waxman}  \& {Portegies
  Zwart}}{{Hopman} et~al.}{2006}]{2006ApJ...643L..91H}
{Hopman} C.,  {Guetta} D.,  {Waxman} E.,   {Portegies Zwart} S.,  2006, \mn@doi
  [\apjl] {10.1086/505141}, \href
  {https://ui.adsabs.harvard.edu/abs/2006ApJ...643L..91H} {643, L91}

\bibitem[\protect\citeauthoryear{{Hotokezaka}, {Kyutoku}, {Okawa}, {Shibata}
  \& {Kiuchi}}{{Hotokezaka} et~al.}{2011}]{2011PhRvD..83l4008H}
{Hotokezaka} K.,  {Kyutoku} K.,  {Okawa} H.,  {Shibata} M.,   {Kiuchi} K.,
  2011, \mn@doi [\prd] {10.1103/PhysRevD.83.124008}, \href
  {https://ui.adsabs.harvard.edu/abs/2011PhRvD..83l4008H} {83, 124008}

\bibitem[\protect\citeauthoryear{{Jonker} et~al.,}{{Jonker}
  et~al.}{2013}]{2013arXiv1306.2336J}
{Jonker} P.,  et~al., 2013, arXiv e-prints, \href
  {https://ui.adsabs.harvard.edu/abs/2013arXiv1306.2336J} {p. arXiv:1306.2336}

\bibitem[\protect\citeauthoryear{{Lattimer} \& {Prakash}}{{Lattimer} \&
  {Prakash}}{2001}]{2001ApJ...550..426L}
{Lattimer} J.~M.,  {Prakash} M.,  2001, \mn@doi [\apj] {10.1086/319702}, \href
  {https://ui.adsabs.harvard.edu/abs/2001ApJ...550..426L} {550, 426}

\bibitem[\protect\citeauthoryear{{L{\"u}} et~al.,}{{L{\"u}}
  et~al.}{2020}]{2020arXiv200701124L}
{L{\"u}} H.-J.,  et~al., 2020, arXiv e-prints, \href
  {https://ui.adsabs.harvard.edu/abs/2020arXiv200701124L} {p. arXiv:2007.01124}

\bibitem[\protect\citeauthoryear{{Mapelli} \& {Giacobbo}}{{Mapelli} \&
  {Giacobbo}}{2018}]{2018MNRAS.479.4391M}
{Mapelli} M.,  {Giacobbo} N.,  2018, \mn@doi [\mnras] {10.1093/mnras/sty1613},
  \href {https://ui.adsabs.harvard.edu/abs/2018MNRAS.479.4391M} {479, 4391}

\bibitem[\protect\citeauthoryear{{Mapelli}, {Giacobbo}, {Santoliquido}  \&
  {Artale}}{{Mapelli} et~al.}{2019}]{2019MNRAS.487....2M}
{Mapelli} M.,  {Giacobbo} N.,  {Santoliquido} F.,   {Artale} M.~C.,  2019,
  \mn@doi [\mnras] {10.1093/mnras/stz1150}, \href
  {https://ui.adsabs.harvard.edu/abs/2019MNRAS.487....2M} {487, 2}

\bibitem[\protect\citeauthoryear{{Margalit} \& {Metzger}}{{Margalit} \&
  {Metzger}}{2017}]{2017ApJ...850L..19M}
{Margalit} B.,  {Metzger} B.~D.,  2017, \mn@doi [\apjl]
  {10.3847/2041-8213/aa991c}, \href
  {https://ui.adsabs.harvard.edu/abs/2017ApJ...850L..19M} {850, L19}

\bibitem[\protect\citeauthoryear{{Martinez} et~al.,}{{Martinez}
  et~al.}{2015}]{2015ApJ...812..143M}
{Martinez} J.~G.,  et~al., 2015, \mn@doi [\apj] {10.1088/0004-637X/812/2/143},
  \href {https://ui.adsabs.harvard.edu/abs/2015ApJ...812..143M} {812, 143}

\bibitem[\protect\citeauthoryear{{Metzger} \& {Piro}}{{Metzger} \&
  {Piro}}{2014}]{2014MNRAS.439.3916M}
{Metzger} B.~D.,  {Piro} A.~L.,  2014, \mn@doi [\mnras] {10.1093/mnras/stu247},
  \href {https://ui.adsabs.harvard.edu/abs/2014MNRAS.439.3916M} {439, 3916}

\bibitem[\protect\citeauthoryear{{Metzger}, {Quataert}  \&
  {Thompson}}{{Metzger} et~al.}{2008}]{2008MNRAS.385.1455M}
{Metzger} B.~D.,  {Quataert} E.,   {Thompson} T.~A.,  2008, \mn@doi [MNRAS]
  {10.1111/j.1365-2966.2008.12923.x}, \href
  {http://adsabs.harvard.edu/abs/2008MNRAS.385.1455M} {385, 1455}

\bibitem[\protect\citeauthoryear{{Metzger}, {Giannios}, {Thompson},
  {Bucciantini}  \& {Quataert}}{{Metzger} et~al.}{2011}]{2011MNRAS.413.2031M}
{Metzger} B.~D.,  {Giannios} D.,  {Thompson} T.~A.,  {Bucciantini} N.,
  {Quataert} E.,  2011, \mn@doi [MNRAS] {10.1111/j.1365-2966.2011.18280.x},
  \href {http://adsabs.harvard.edu/abs/2011MNRAS.413.2031M} {413, 2031}

\bibitem[\protect\citeauthoryear{{M{\"u}ller} \& {Serot}}{{M{\"u}ller} \&
  {Serot}}{1996}]{1996NuPhA.606..508M}
{M{\"u}ller} H.,  {Serot} B.~D.,  1996, \mn@doi [\nphysa]
  {10.1016/0375-9474(96)00187-X}, \href
  {https://ui.adsabs.harvard.edu/abs/1996NuPhA.606..508M} {606, 508}

\bibitem[\protect\citeauthoryear{{Nakar}, {Gal-Yam}  \& {Fox}}{{Nakar}
  et~al.}{2006}]{2006ApJ...650..281N}
{Nakar} E.,  {Gal-Yam} A.,   {Fox} D.~B.,  2006, \mn@doi [\apj]
  {10.1086/505855}, \href
  {https://ui.adsabs.harvard.edu/abs/2006ApJ...650..281N} {650, 281}

\bibitem[\protect\citeauthoryear{{Norris} \& {Bonnell}}{{Norris} \&
  {Bonnell}}{2006}]{2006ApJ...643..266N}
{Norris} J.~P.,  {Bonnell} J.~T.,  2006, \mn@doi [ApJ] {10.1086/502796}, \href
  {http://adsabs.harvard.edu/abs/2006ApJ...643..266N} {643, 266}

\bibitem[\protect\citeauthoryear{{Oganesyan}, {Ascenzi}, {Branchesi},
  {Salafia}, {Dall'Osso}  \& {Ghirlanda}}{{Oganesyan}
  et~al.}{2020}]{2020ApJ...893...88O}
{Oganesyan} G.,  {Ascenzi} S.,  {Branchesi} M.,  {Salafia} O.~S.,  {Dall'Osso}
  S.,   {Ghirlanda} G.,  2020, \mn@doi [\apj] {10.3847/1538-4357/ab8221}, \href
  {https://ui.adsabs.harvard.edu/abs/2020ApJ...893...88O} {893, 88}

\bibitem[\protect\citeauthoryear{{{\"O}zel} \& {Freire}}{{{\"O}zel} \&
  {Freire}}{2016}]{2016ARA&A..54..401O}
{{\"O}zel} F.,  {Freire} P.,  2016, \mn@doi [\araa]
  {10.1146/annurev-astro-081915-023322}, \href
  {https://ui.adsabs.harvard.edu/abs/2016ARA&A..54..401O} {54, 401}

\bibitem[\protect\citeauthoryear{{Piro}, {Giacomazzo}  \& {Perna}}{{Piro}
  et~al.}{2017}]{2017ApJ...844L..19P}
{Piro} A.~L.,  {Giacomazzo} B.,   {Perna} R.,  2017, \mn@doi [\apjl]
  {10.3847/2041-8213/aa7f2f}, \href
  {https://ui.adsabs.harvard.edu/abs/2017ApJ...844L..19P} {844, L19}

\bibitem[\protect\citeauthoryear{{Piro} et~al.,}{{Piro}
  et~al.}{2019}]{2019MNRAS.483.1912P}
{Piro} L.,  et~al., 2019, \mn@doi [\mnras] {10.1093/mnras/sty3047}, \href
  {https://ui.adsabs.harvard.edu/abs/2019MNRAS.483.1912P} {483, 1912}

\bibitem[\protect\citeauthoryear{{Raaijmakers} et~al.,}{{Raaijmakers}
  et~al.}{2020}]{2020ApJ...893L..21R}
{Raaijmakers} G.,  et~al., 2020, \mn@doi [\apjl] {10.3847/2041-8213/ab822f},
  \href {https://ui.adsabs.harvard.edu/abs/2020ApJ...893L..21R} {893, L21}

\bibitem[\protect\citeauthoryear{{Radice}, {Perego}, {Zappa}  \&
  {Bernuzzi}}{{Radice} et~al.}{2018}]{2018ApJ...852L..29R}
{Radice} D.,  {Perego} A.,  {Zappa} F.,   {Bernuzzi} S.,  2018, \mn@doi [\apjl]
  {10.3847/2041-8213/aaa402}, \href
  {https://ui.adsabs.harvard.edu/abs/2018ApJ...852L..29R} {852, L29}

\bibitem[\protect\citeauthoryear{{Rowlinson}, {O'Brien}, {Metzger}, {Tanvir}
  \& {Levan}}{{Rowlinson} et~al.}{2013}]{2013MNRAS.430.1061R}
{Rowlinson} A.,  {O'Brien} P.~T.,  {Metzger} B.~D.,  {Tanvir} N.~R.,   {Levan}
  A.~J.,  2013, \mn@doi [MNRAS] {10.1093/mnras/sts683}, \href
  {http://adsabs.harvard.edu/abs/2013MNRAS.430.1061R} {430, 1061}

\bibitem[\protect\citeauthoryear{{Safarzadeh}, {Ramirez-Ruiz}  \&
  {Berger}}{{Safarzadeh} et~al.}{2020}]{2020ApJ...900...13S}
{Safarzadeh} M.,  {Ramirez-Ruiz} E.,   {Berger} E.,  2020, \mn@doi [\apj]
  {10.3847/1538-4357/aba596}, \href
  {https://ui.adsabs.harvard.edu/abs/2020ApJ...900...13S} {900, 13}

\bibitem[\protect\citeauthoryear{{Salafia}, {Ghisellini}, {Pescalli}, {Ghirland
  a}  \& {Nappo}}{{Salafia} et~al.}{2015}]{2015MNRAS.450.3549S}
{Salafia} O.~S.,  {Ghisellini} G.,  {Pescalli} A.,  {Ghirland a} G.,   {Nappo}
  F.,  2015, \mn@doi [\mnras] {10.1093/mnras/stv766}, \href
  {https://ui.adsabs.harvard.edu/abs/2015MNRAS.450.3549S} {450, 3549}

\bibitem[\protect\citeauthoryear{{Salafia}, {Barbieri}, {Ascenzi}  \&
  {Toffano}}{{Salafia} et~al.}{2020}]{2020A&A...636A.105S}
{Salafia} O.~S.,  {Barbieri} C.,  {Ascenzi} S.,   {Toffano} M.,  2020, \mn@doi
  [\aap] {10.1051/0004-6361/201936335}, \href
  {https://ui.adsabs.harvard.edu/abs/2020A&A...636A.105S} {636, A105}

\bibitem[\protect\citeauthoryear{{Salvaterra}, {Cerutti}, {Chincarini},
  {Colpi}, {Guidorzi}  \& {Romano}}{{Salvaterra}
  et~al.}{2008}]{2008MNRAS.388L...6S}
{Salvaterra} R.,  {Cerutti} A.,  {Chincarini} G.,  {Colpi} M.,  {Guidorzi} C.,
   {Romano} P.,  2008, \mn@doi [\mnras] {10.1111/j.1745-3933.2008.00488.x},
  \href {https://ui.adsabs.harvard.edu/abs/2008MNRAS.388L...6S} {388, L6}

\bibitem[\protect\citeauthoryear{{Santoliquido}, {Mapelli}, {Bouffanais},
  {Giacobbo}, {Di Carlo}, {Rastello}, {Artale}  \& {Ballone}}{{Santoliquido}
  et~al.}{2020}]{2020arXiv200409533S}
{Santoliquido} F.,  {Mapelli} M.,  {Bouffanais} Y.,  {Giacobbo} N.,  {Di Carlo}
  U.~N.,  {Rastello} S.,  {Artale} M.~C.,   {Ballone} A.,  2020, arXiv
  e-prints, \href {https://ui.adsabs.harvard.edu/abs/2020arXiv200409533S} {p.
  arXiv:2004.09533}

\bibitem[\protect\citeauthoryear{{Savchenko} et~al.,}{{Savchenko}
  et~al.}{2017}]{2017ApJ...848L..15S}
{Savchenko} V.,  et~al., 2017, \mn@doi [\apjl] {10.3847/2041-8213/aa8f94},
  \href {https://ui.adsabs.harvard.edu/abs/2017ApJ...848L..15S} {848, L15}

\bibitem[\protect\citeauthoryear{{Shibata} \& {Taniguchi}}{{Shibata} \&
  {Taniguchi}}{2006}]{2006PhRvD..73f4027S}
{Shibata} M.,  {Taniguchi} K.,  2006, \mn@doi [\prd]
  {10.1103/PhysRevD.73.064027}, \href
  {https://ui.adsabs.harvard.edu/abs/2006PhRvD..73f4027S} {73, 064027}

\bibitem[\protect\citeauthoryear{{Shibata}, {Fujibayashi}, {Hotokezaka},
  {Kiuchi}, {Kyutoku}, {Sekiguchi}  \& {Tanaka}}{{Shibata}
  et~al.}{2017}]{2017PhRvD..96l3012S}
{Shibata} M.,  {Fujibayashi} S.,  {Hotokezaka} K.,  {Kiuchi} K.,  {Kyutoku} K.,
   {Sekiguchi} Y.,   {Tanaka} M.,  2017, \mn@doi [\prd]
  {10.1103/PhysRevD.96.123012}, \href
  {https://ui.adsabs.harvard.edu/abs/2017PhRvD..96l3012S} {96, 123012}

\bibitem[\protect\citeauthoryear{{Siegel} \& {Ciolfi}}{{Siegel} \&
  {Ciolfi}}{2016a}]{2016ApJ...819...14S}
{Siegel} D.~M.,  {Ciolfi} R.,  2016a, \mn@doi [\apj]
  {10.3847/0004-637X/819/1/14}, \href
  {https://ui.adsabs.harvard.edu/abs/2016ApJ...819...14S} {819, 14}

\bibitem[\protect\citeauthoryear{{Siegel} \& {Ciolfi}}{{Siegel} \&
  {Ciolfi}}{2016b}]{2016ApJ...819...15S}
{Siegel} D.~M.,  {Ciolfi} R.,  2016b, \mn@doi [\apj]
  {10.3847/0004-637X/819/1/15}, \href
  {https://ui.adsabs.harvard.edu/abs/2016ApJ...819...15S} {819, 15}

\bibitem[\protect\citeauthoryear{{Troja}, {King}, {O'Brien}, {Lyons}  \&
  {Cusumano}}{{Troja} et~al.}{2008}]{2008MNRAS.385L..10T}
{Troja} E.,  {King} A.~R.,  {O'Brien} P.~T.,  {Lyons} N.,   {Cusumano} G.,
  2008, \mn@doi [\mnras] {10.1111/j.1745-3933.2007.00421.x}, \href
  {https://ui.adsabs.harvard.edu/abs/2008MNRAS.385L..10T} {385, L10}

\bibitem[\protect\citeauthoryear{{Virgili}, {Zhang}, {O'Brien}  \&
  {Troja}}{{Virgili} et~al.}{2011}]{2011ApJ...727..109V}
{Virgili} F.~J.,  {Zhang} B.,  {O'Brien} P.,   {Troja} E.,  2011, \mn@doi
  [\apj] {10.1088/0004-637X/727/2/109}, \href
  {https://ui.adsabs.harvard.edu/abs/2011ApJ...727..109V} {727, 109}

\bibitem[\protect\citeauthoryear{{Wanderman} \& {Piran}}{{Wanderman} \&
  {Piran}}{2015}]{2015MNRAS.448.3026W}
{Wanderman} D.,  {Piran} T.,  2015, \mn@doi [\mnras] {10.1093/mnras/stv123},
  \href {https://ui.adsabs.harvard.edu/abs/2015MNRAS.448.3026W} {448, 3026}

\bibitem[\protect\citeauthoryear{{Ye}, {Fong}, {Kremer}, {Rodriguez},
  {Chatterjee}, {Fragione}  \& {Rasio}}{{Ye}
  et~al.}{2020}]{2020ApJ...888L..10Y}
{Ye} C.~S.,  {Fong} W.-f.,  {Kremer} K.,  {Rodriguez} C.~L.,  {Chatterjee} S.,
  {Fragione} G.,   {Rasio} F.~A.,  2020, \mn@doi [\apjl]
  {10.3847/2041-8213/ab5dc5}, \href
  {https://ui.adsabs.harvard.edu/abs/2020ApJ...888L..10Y} {888, L10}

\bibitem[\protect\citeauthoryear{{Yu}, {Zhang}  \& {Gao}}{{Yu}
  et~al.}{2013}]{2013ApJ...776L..40Y}
{Yu} Y.-W.,  {Zhang} B.,   {Gao} H.,  2013, \mn@doi [\apjl]
  {10.1088/2041-8205/776/2/L40}, \href
  {https://ui.adsabs.harvard.edu/abs/2013ApJ...776L..40Y} {776, L40}

\bibitem[\protect\citeauthoryear{{Zhang} \& {M{\'e}sz{\'a}ros}}{{Zhang} \&
  {M{\'e}sz{\'a}ros}}{2001}]{2001ApJ...552L..35Z}
{Zhang} B.,  {M{\'e}sz{\'a}ros} P.,  2001, \mn@doi [ApJ] {10.1086/320255},
  \href {http://adsabs.harvard.edu/abs/2001ApJ...552L..35Z} {552, L35}

\makeatother
\end{thebibliography}








\bsp	
\label{lastpage}
\end{document}